\documentclass[12pt]{article}

\usepackage[utf8]{inputenc}
\usepackage{charter}
\usepackage{fullpage}
\usepackage{hyperref}
\usepackage{graphicx}
\usepackage{amssymb,amsmath}
\usepackage{lipsum}
\usepackage[sort&compress,numbers]{natbib}
\usepackage{hyperref}
\hypersetup{hidelinks}
\usepackage{wrapfig}
\usepackage[usenames,dvipsnames,table]{xcolor}
\usepackage[total={6.5in,9in},top=1in,headsep=0.1in,headheight=1in]{geometry}
\usepackage{lineno}

\newcommand\snowmass{
\begin{center}
  \rule[-0.2in]{\hsize}{0.01in}\\
  \rule{\hsize}{0.01in}\\
  \vskip 0.1in
  Submitted to the Proceedings of the US Community Study\\ 
  on the Future of Particle Physics (Snowmass 2021)\\
  \rule{\hsize}{0.01in}\\
  \rule[+0.2in]{\hsize}{0.01in}\\[-2em]
\end{center}
}
 
 \usepackage[firstpage=true]{background}
\backgroundsetup{contents={\parbox{6.5in}{\snowmass}}, scale=1,placement=top,opacity=1,color=black,position={3.25in,1.2in}}

\usepackage{fancyhdr}
\fancypagestyle{plain}{%
  \fancyhf{}%
  \fancyhead[C]{}
  \fancyfoot[C]{\thepage}
}

\fancypagestyle{empty}{%
  \fancyhf{}%
  \fancyhead[C]{{\it Snowmass White Paper: AI and Uncertainty Quantification}}
  \fancyfoot[C]{\thepage}
}
\usepackage[utf8]{inputenc}
\usepackage{geometry}
\usepackage{fullpage}
\usepackage[T1]{fontenc}    
\usepackage{hyperref} 
\usepackage{bm}	
\usepackage[normalem]{ulem}
\usepackage{caption}
\usepackage{subcaption}
\usepackage{verbatim}
\usepackage{lineno}
\usepackage{url}    
\usepackage{booktabs}       
\usepackage{amsfonts}       
\usepackage{nicefrac}
\usepackage{natbib}
\usepackage{xcolor}

\usepackage{authblk}

\author[1]{Thomas Y. Chen} 
\affil[1]{Fu Foundation School of Engineering and Applied Science, Columbia University, New York, NY 10027, USA}
\author[2]{Biprateep Dey}
\affil[2]{Department of Physics and Astronomy and PITT PACC, University of Pittsburgh, Pittsburgh, PA 15260, USA}
\author[3,4]{Aishik Ghosh}
\affil[3]{Department of Physics and Astronomy, University of California, Irvine, CA 92697, USA}
\affil[4]{Physics Division, Lawrence Berkeley National Laboratory, Berkeley, CA 94720, USA}
\author[5]{Michael Kagan}
\affil[5]{SLAC National Accelerator Laboratory, Menlo Park, CA 94025, USA}
\author[6,7,8]{Brian Nord}
\affil[6]{Fermi National Accelerator Laboratory, P.O. Box 500, Batavia, IL 60510, USA}
\affil[7]{Kavli Institute for Cosmological Physics, University of Chicago, Chicago, IL 60637, USA}
\affil[8]{Department of Astronomy and Astrophysics, University of Chicago, Chicago, IL 60637, USA}
\author[9]{Nesar Ramachandra}
\affil[9]{Computational Science Division, Argonne National Laboratory, 9700 South Cass Avenue, Lemont, IL 60439, USA}

\title{Interpretable Uncertainty Quantification \\ in AI for HEP}

\begin{document}

\maketitle

\begin{abstract}
Estimating uncertainty is at the core of performing scientific measurements in HEP: a measurement is not useful without an estimate of its uncertainty. 
The goal of uncertainty quantification (UQ) is inextricably linked to the question, ``how do we physically and statistically interpret these uncertainties?''
The answer to this question depends not only on the computational task we aim to undertake, but also on the methods we use for that task.
For artificial intelligence (AI) applications in HEP, there are several areas where interpretable methods for UQ are essential, including inference, simulation, and control/decision-making. 
There exist some methods for each of these areas, but they have not yet been demonstrated to be as trustworthy as more traditional approaches currently employed in physics (e.g., non-AI frequentist and Bayesian methods).

Shedding light on the questions above requires additional understanding of the interplay of AI systems and uncertainty quantification. 
We briefly discuss the existing methods in each area and relate them to tasks across HEP. 
We then discuss recommendations for avenues to pursue to develop the necessary techniques for reliable widespread usage of AI with UQ over the next decade.

\end{abstract}




\section{Requirements for uncertainties in physics}

What are the uncertainties we consider in HEP and in AI, and how do they relate or correspond to each other? 
The following brief text benefits from the excellent discussion on this topic in 2021 PDG review of AI in HEP~\cite{Zyla:2020zbs}.

\subsection{Definitions of Uncertainty in HEP}
In HEP, uncertainties are often categorized as ``statistical'' and ``systematic.'' 
Statistical uncertainties typically refer to error associated with the finite sizes of a data set. For instance, in collider physics often a feature of the data, whose distribution is sensitive to a parameter of interest, is computed for each collision event and the population is binned for statistical inference. As such, the statistical model of counts in each bin follows a Poisson distribution (e.g. the bins have a Poisson statistical error). Similarly in Astrophysics, photon counting in measure image pixels follows a similar Poisson distribution.

Systematic uncertainties typically arise due to an incomplete knowledge of the true data generation process.
When simulations (forward models) or analytic expressions (inverse models) are constructed with this incomplete knowledge, the difference is propagated through inference process to create a systematic error.
For example, this could be uncertainty in the theoretical model of particle scattering, in particle-material interactions, detector calibrations, or in the material description of a given detector.  
Systematic uncertainties can also arise during the use of data-driven models, like those used in AI.
Consider, for example, in the case where control samples of the data  are selected and used to form a model of a background process that mimics a signal. 
Systematic uncertainties are often described with nuisance parameter $\boldsymbol{\nu}$, which  parameterizes a continuous family of data-generating processes.
Predictions depend on these nuisance parameters. 
When simulations and control samples don't match the original data (which is always), we encounter a domain transition (or adaptation) challenge

In addition to statistical and systematic uncertainties, the process of detecting particles is inherently uncertain and detectors have finite resolution; this resolution can be considered an inherent uncertainty in the detection process.

\subsection{Definitions of Uncertainty in AI}
In AI, uncertainty is often categorized broadly as ``aleatoric'' and ``epistemic.'' 
``Aleatoric'' uncertainty is often noted as being due to inherent randomness in the outcome of an experiment. 
``Epistemic'' uncertainty is often described as coming from a lack of knowledge -- e.g. a lack of knowledge about the model. 
In the AI literature, aleatoric uncertainty is often considered irreducible and coming from the data generation process, while epistemic uncertainty is discussed in terms of the uncertainty in model parameters and reduces when more data is added to the training data set.

Statistical error due to finite sample sizes and inherent randomness in measurements -- e.g. due to finite detector resolutions -- are concepts well studied in HEP. 
Model uncertainty in HEP often takes the form of a lack of knowledge in our physics models and is irreducible without further measurements~\cite{Zyla:2020zbs}, and thus is more akin to a systematic uncertainty, like a domain shift. 
More generally, the usage of ML uncertainty terminology and their correspondence to HEP uncertainty terminology is often ambiguous. 
How we use and interpret the various kinds uncertainties in a consistent way across AI-related analyses in HEP remains an unresolved discussion and problem.

The use of AI is growing across many sectors of HEP -- e.g., reconstructing detector data for estimating physical objects like particle momenta or galaxy properties; classifying detection events as signal or background;  estimating the density of populations of events;  defining low-dimensional powerful features that can be used for inference; constructing surrogate models of large-scale simulations; scheduling experimental data acquisition and operating instruments. 

Depending on the task ML is used for and how inference on a data set is performed, different kinds of uncertainties may be needed. In cases where the ML model predictions determine the quality of a statistical model used to compare to data, careful attention to uncertainty is needed. In other cases the quality of the ML model may determine the optimality of an analysis or test statistic, but would would not lead to incorrect inference results; if and what uncertainties are needed in such cases also requires careful consideration.
In the case of problems relating to instrument/experiment control, where the goal is predictive modeling of future states and  to determine actions for a system to take, the full suite of related systematic uncertainties -- AI model parameter uncertainty, resolutions of measurements, statistical, and domain shift -- are notably essential to assess risk of and protect the data acquisition process and the instrument itself.

\subsection{Inference for Parameter Estimation}

A primary goal in HEP is to estimate the parameters $\boldsymbol{\theta}$ of a physical model -- e.g., the Standard Model of particle physics or the cosmological $\Lambda$CDM model -- from experimental data, $\boldsymbol{y}$.  
In collider physics, frequentist methods are common and the profile likelihood ratio test statistic is typically used for confidence interval estimation. In addition, in collider physics inference often takes the form of unfolding, wherein the corruption of particle properties due to the stochastic nature of detectors is removed (e.g. de-convolved). 
Bayesian inference has been in widespread use in cosmology since approximately 2010 -- almost to the exception of frequentism.
Indeed, the field is moving toward the use of Bayesian hierarchical models to simultaneous capture global features of the universe (e.g., cosmic dark matter density) and local features of individual objects (e.g., mass of one galaxy).

When one aims to model high-dimensional feature spaces, the estimation of the likelihood, the likelihood ratio, and the posterior  all become extremely challenging. 
Simulation-Based Inference (SBI; aka, likelihood-free inference or implicit likelihood inference) has emerged as a subfield that uses AI with multi-fidelity simulators to efficiently and effectively perform such inference tasks~\cite{Cranmer:2019eaq}.  
Using the likelihood ratio trick, learned classifiers between different parameter hypothesis can be used to approximate likelihood ratios, and several strategies to perform this estimation efficiently have been proposed~\cite{pnas.1915980117}. 
Similarly, using powerful normalizing flow models, which are capable of explicitly modeling high dimensional data densities, schemes have been developed to estimate likelihoods and posteriors~\cite{JMLR:v22:19-1028}.

\section{Modern Approaches to UQ in AI}

Predictive uncertainty associated with the output depends on both the AI model and the data. 
The terms ``epistemic uncertainty'' (also referred to as ``systematic'' or  ``model'' uncertainty) and   ``aleatoric uncertainty'' (also known as  ``statistical'' or  ``data'' uncertainty) are commonly used in AI literature to differentiate these sources of errors \cite{hullermeier:2021aleatoric, der2009aleatory}.
However, while the terms are sometimes used interchangeably, these correspondences don't consistently hold up across all problems and model-building scenarios.

In the context of scientific AI, however, the predictive uncertainties also depend on the mismatch between the data domains. 
For instance, a deep neural network (DNN) trained solely on numerical astrophysical simulations will have additional uncertainties associated when applied to real telescope observations. 
Such dataset-dependent uncertainties could be due to a domain transition (when the coverage or distribution of training data does not match with those of the real data) or due to out-of-domain information (training simulations not capturing all the effects of the physical phenomena) \cite{gawlikowski:2021survey}. 
The above classifications are neither universal nor exhaustive, and uncertainty quantification must be specific to a particular AI approach to the scientific problem. 
The exact representation of uncertainty in an AI approach is often not fully known, due to the caveats mentioned above. 
Memory usage, convergence or training time, scaling across computing nodes, and other practical considerations play vital roles in the particular choice of a model for a given problem.  
Hence, AI-based uncertainty approximation remains  a popular research avenue with variety of methods in active development \cite{li:2012dealing, gawlikowski:2021survey, staahl:2020evaluation}. 

In this section, we briefly mention a few popular categories of these UQ approaches. 

\begin{itemize}
    \item {\bf Emulator-based posterior estimation:} 
 A conservative approach to performing Bayesian approximation involves using AI surrogates in forward models or explicit likelihoods in Bayesian inference schemes such as Markhov Chain Monte Carlo (MCMC) techniques. 
 These emulator-based approaches \citep[e.g.,][]{heitmann2013coyote, McClintock2019,Zhai2019,Kobayashi2020, ramachandra2021matter} are now widely used in cosmological survey sciences. 
 Coupled with active sampling techniques, one may also consider SBI methods \cite{alsing2019fast} using neural density estimators. 
 Recent work on differentiable emulators \citep{modi2021flowpm, hearin2021differentiable, hearin2021differentiable1} also paves the way for faster posterior sampling via Hamiltonian Monte Carlo (HMC) techniques. 

\item {\bf Sampling over network parameters:} neural networks could be used as inference machines as alternatives to traditional posterior estimation techniques, by sampling over the model parameters of network weights and biases \citep{neal2012bayesian}. 
This involves treating neural network's parameters as something you'd want to do posterior analysis over, by MCMC (or some other technique).
The numerical expense of sampling over a large number of model parameters (multiple training experiments?) can be numerically expensive, as is the case with deep neural networks. 
However, with GPU-accelerators and AI-specific hardware, HMC sampling over a relatively small number network parameters will be within reach in the near future. 

\item {\bf Variational Inference (VI):} This family of approaches approximate and learn the posterior distribution over network weights and biases. These VI-based methods effectively convert the statistical inference problem to a familiar network optimization problem, where a loss metric is optimized via gradient descent algorithms. 
A popular network that uses VI is the variational auto-encoder used in HEP problems that require generative modeling or dimension-reduction \cite{cerri2019variational, larkoski2020jet}. 
Normalizing flow models \cite{rezende2015variational} also use VI to construct complex distributions by transforming a probability density through a series of invertible mappings. 
Another widely used VI approach of Monte Carlo Dropout (MC Dropout) uses a regularization technique of dropout layers \cite{gal2016dropout} to an approximate a variational distribution. 
The trade-off between the quality of uncertainty estimates and the computational complexity is a notable feature of various VI-approaches: versions of MC dropout provice uncertainty estimates with little to no overhead during training time. 

\item {\bf Deep ensemble methods:} Another approach to predicting UQ from point-estimating DNNs is to train a large number of networks, and use the ensemble distribution to predict the posterior prediction \cite{DeepEnsembles2016}. 
With an ensemble, each individually trained network has its own loss function. 
This is in contrast to sampling over NN's, where there aren't individual loss functions
The ensemble of networks can be obtained by either on batches of data or different architectures, random weight initializations, MC Dropout, bootstrapping or other techniques. 
Ensemble-based UQ using sampling techniques like active learning or hyperparameter optimization \cite{balaprakash2018deephyper, egele2021autodeuq} are also currently explored in massive supercomputing environments. 
Averaging in the weight space, instead of the predictions, at different stages of training also results in a computationally inexpensive and reliable estimate of UQ \cite{izmailov2018averaging}. 

\item {\bf Deep Gaussian Processes:} A Conventional Gaussian process (GP) is a non-parametric Bayesian model that provides uncertainty estimates on predictions, even with small number of training datapoints. 
The Deep GP is a composition of GPs that can effectively scale at large datasets using VI \cite{damianou2013deep}. 
Deep GPs are demonstrated to perform better than several AI UQ techniques for domain shifts and against adversarial agents \cite{bradshaw2017adversarial}.

\item {\bf Test-time implementations:} Multiple predictions from a point-prediction or a probabilistic network can be augmented to evaluate the error in the predictions \cite{simonyan2014very}. 
Such approaches are extremely useful in error propagation, wherein the input data from the sensors can be noisy \cite{ramachandra2021machine}. 

\item {\bf Bayesian Neural Networks (BNNs):} These are typically optimized through VI.
These incorporate parameters of a distribution for the the weights of an activation neuron~\cite{Gal2016Uncertainty}.

\item {\bf Conformal methods:} A class of algorithms that use past experiences to determine levels of confidence in new predictions. 
Once a point prediction is made, a non-conformity score is constructed, which is a measure of how unusual the new example is compared to previously seen examples -- i.e., a labeled calibration set. 
The conformal algorithm then converts this non-conformity score into a prediction region. 
Conformal methods can be used with any estimator, are computationally efficient and guarantee finite sample marginal coverage \cite{shafer2008Conformal,lei2018Conformal}. 

\item {\bf Simulation-based inference (SBI):} Also known as implicit likelihood inference (or the misnomer, likelihood-free inference), SBI sequesters the model to the simulator only \citep{Cranmer:2019eaq}. 
For a given choice of parameters of interest, a high-fidelity simulation (including the internal/hidden physical mechanisms and the observational signatures) is compared to observations. 
If the simulation matches the observation, these parameters are selected as the best parameters to represent reality. 
Sometimes, these methods use normalizing flows for neural density estimation.

There are also methods in Reinforcement Learning and other areas of ML that are of interest to physics, which may be used in observation scheduling and instrument control. 
Some of these are discussed in this recent review \cite{Abdar_2021}.

\end{itemize}

Numerous implementations of these UQ algorithm families  have started to shift the landscape of AI from deterministic models to those that are more probabilistic.
Depending on the research problem, each of these methods have advantages and drawbacks. 
For instance, the MC dropout and test-time approaches are computationally inexpensive compared to the methods that involve sampling. 
We restrict the discussion of the UQ methods to the above, but note again that numerous implementations within each family tackle the issue of achieving robust UQ predictions using neural networks.

\subsection{Calibration of Uncertainty Estimates}
AI-based algorithms quote uncertainty estimates as probability distributions: they may be class probabilities for a classification task, or a credibility interval or a full probability distribution function for regression tasks. Calibration refers to the property of such probabilistic estimates to be consistent with the frequency of occurrence of the target variable in observed populations. 
For example, if an algorithm predicts a class probability of 0.8 or produces an 80\% credibility interval, then the true value should fall in the predicted class or the credibility interval 80\% of the time for any arbitrary test set. Though calibration seems like a simple property, most common machine learning algorithms provide mis-calibrated probabilistic estimates~\cite{NiculescuEtAl2005Calibration,Guo2017Calibration,Kompa2021Calibration,MalininG18Miscalibration, Lakshminarayanan17Miscalibration}.
Though the term ``calibration'' is widely used to refer to this property in the AI community, the term ``coverage'' is more popular in the statistics community.

We want models to produce calibrated uncertainty estimates, because we would like to interpret probabilistic estimates as long run frequencies. 
Mis-calibrated uncertainty estimates used as input for a physical analysis will result in biased estimates of physical quantities.

Calibration performance of AI-based methods is generally assessed by comparing the predicted probability of a prediction belonging to a class (or a credibility interval) against their empirical probabilities calculated using a test set: ideally both the quantities should be equal. 
This is quantified using metrics like Expected Calibration Error \cite[ECE;][]{Naeini2015ECE}, Brier score~\cite{Brier1950BrierScore}, Probability Integral Transform~\cite[PIT;][]{Gan90PIT}, etc. 
All of these tests fail to condition on the class and input features, but rather asses the calibration of probabilistic estimates as an ensemble and not on an individual basis \cite{nixon19calibration,zhao21Calibration}. 
Moreover some of these metrics are optimized through completely un-physical probability estimates\cite{schmidt2020}. Therefore, metrics and tests should be developed to measure the calibration of individual uncertainty estimates (i.e., local calibration/conditional coverage) and not just for ensembles (i.e., global calibration/marginal coverage) as inconsistencies in various regions of the feature space cancel out to produce deceptively optimal results when looked as an ensemble\cite{zhao21Calibration,Jitkrittum20KernelStein,Luo2021}.

Generally, it is easier to use generic re-calibration methods on outputs of existing algorithms than to develop custom workarounds for each of them. 
Therefore, methods like Plat scaling~\cite{Platt99probabilisticoutputs} and Isotonic regression~\cite{Zadrozny02Isotonic} are widely used to re-calibrate outputs of AI-based uncertainty estimates.
However, they suffer from the same problems as the metrics discussed immediately above in terms of their inability to ensure local calibration (conditional coverage) and therefore new methods should be developed to perform re-calibration of uncertainty estimates while ensuring local calibration~\cite{Dey2021Recalibration,ding21,dalmasso2021LF2I,dey2022calibrated}.

\subsection{Bias Mitigation}
It is often very expensive to get labeled data from experiments and observations that also span the entire range of feature space of interest, but it is often that we have a complete set of unlabeled or simulated data. One approach to minimizing biases in our training data would be to train on simulations and then use transfer learning (or domain adaptation) methods on observed data. Similarly, we can also use self-supervised learning on an unlabeled data set and then fine tune to labeled data. 

HEP heavily relies on comparing unlabeled (observed) data to labeled (simulation or observed) data) to infer parameters of mathematical models that describe our universe: e.g., AI models are often trained on simulations and then applied to data. 
Systematic differences between the simulations and observed data can thus bias physics results. 
The origin of such differences could be due to limitations in computational power, the limited power of some statistical technique applied, our ignorance about the exact state of a detector, or our inability to correctly model certain physics processes accurately within the simulation. 
The nature of the systematic uncertainty informs the appropriate mitigation/quantification mechanism~\cite{Ramachandra2022}.

\subsubsection{Transfer Learning (TL) and Domain Adaptation (DA)}
When an AI algorithm is trained on data that is not representative of, or biased in relation to, the testing data, this results in differences in the distribution between training and testing sets. 
Training on unbiased data is often not possible (or it is computationally burdensome) due to limitations in generation of the data. 
Domain adaptation and transfer learning are nascent research directions under the umbrella of AI that focus on resolving these differences, aiding in generalization of a HEP events classifier from a source dataset to a target domain. 
For example, TL has been used to fine-tune a DNN trained on generic imagery to suit the particular task of neutrino interaction classification, as an insufficient number of simulated events were available as data to train the specific use-case model directly \cite{chappell2022application}.
Additionally, an adversarial learning-based DA approach was proposed to successfully mitigate bias in simulated event classification at the LHC \cite{NIPS2017_48ab2f9b,clavijo2021adversarial}.

\subsubsection{De-correlation methods to reduce bias in a statistical technique}
A frequently used technique in HEP is to fit a smoothly falling parametric function to a background distribution (in some feature, like mass or momentum) to estimate its contribution in the signal region. 
While applying AI models to filter away uninteresting events improves the sensitivity -- therefore shrinking the statistical uncertainty on the final result -- it can reshape the background distribution in undesired ways, resulting in larger systematic uncertainties.
If the background distribution is no longer smoothly falling, it could result in a large bias.
Several recently proposed AI methods are able to reduce the correlation between the output of the model and some given feature to avoid sculpting the background distribution, or between the output of the model and nuisance parameters defining systematic uncertainties~\cite{NIPS2017_48ab2f9b,PhysRevLett.125.122001}. 
This sacrifices classification power to reduce the overall bias, but therefore improves the total (statistical and systematic) uncertainty for the final physics result.

\subsubsection{Uncertainties with a statistical origin}
Some systematic uncertainties have a statistical origin, for example the state of a detector, $\boldsymbol{z}$, may be measured or calibrated using some auxiliary data and reported as a maximum likelihood estimate, $z_0$, along with an uncertainty. 
This uncertainty provides a probabilistic prior $P(\boldsymbol{z})$ on $z$. 
If an AI model is trained on simulations that assume the detector is in state $z_0$, the uncertainty can then be estimated by evaluating the bias of the model on other simulations that assume the detector in some other states, $z_-, z_+$. 
The final physics result may be reported such that the total uncertainty accounts for every possible bias in $\boldsymbol{z}$, as well as its associated prior in $P(\boldsymbol{z})$.

AI offers an opportunity to take it a step further by  reducing the final uncertainty of such a physics result. 
De-correlation methods may also be applied in this case to make the AI model insensitive to change in $\boldsymbol{z}$, again sacrificing the classification power of the model in favor invariance to $\boldsymbol{z}$, which results in increased statistical uncertainties but reduces systematic uncertainties. 
For example, a classifier can be forced to perform similarly on simulated and observed data in the control region to ensure that it does not learn any artifacts arising from inaccuracies in the simulation. 
However, for these kinds of uncertainties, a more appropriate solution is precisely the opposite idea, adaptive risk minimisation. 
An uncertainty-aware model is made maximally sensitive to the origin of the bias by parameterizing it on $\boldsymbol{z}$. 
This model provides an optimal classifier for every value of $\boldsymbol{z}$. 
This reduces the statistical component of the final uncertainty and allows us to better extract information about the true value of $\boldsymbol{z}$ from data, which can be combined with the prior $P(\boldsymbol{z})$ to further improve the final result.

While classifiers are trained for all the above discussed methods, the classifier objective acts only as a proxy for the final objective of a physics analysis. 
Inference-aware learning, where the true objective is directly optimized with differentiable programming, provides another tool to directly reduce the total uncertainty of a physics result. 
A model that is both inference-aware and uncertainty-aware is likely to provide close to optimal results. 
Since differentiable programming requires encoding the analysis-specific objective into differentiable code, this is a new skill that physicists will have to acquire in the future.

\subsubsection{Uncertainties without a statistical origin}

In contrast to the kind of uncertainties discussed above, certain systematic uncertainties lack a statistical origin. There is no probabilistic prior and it is often impossible to simulate the full range of possible biases coming from them. For example consider theoretical uncertainties that arise due to our inability to correctly model and compute QFT. 
Different hadronization models~\cite{Sjostrand:2007gs, Bahr:2008pv} predict slightly different results but do not span the full range of possible mismodelling errors. 
The differences between two available hadronization models are often used as an ad hoc quantification of the uncertainty on hadronization modelling. 
The systematic uncertainties that arise because we are unable to compute QFT at infinite order are also estimated using similar well-motivated guesswork. 
Unphysical scales in the computation are varied in the simulation and the sensitivity of a physics result to these variations is treated as a systematic uncertainty. While this helps ascertain a rough estimate of the potential bias, the uncertainty band does not come from any rigorous statistical method for estimating an uncertainty.  

While it may be tempting to apply AI-based de-correlation techniques to make a model insensitive to these biases, there is a danger in this case of significantly underestimating the true bias and uncertainties as discussed in Ref.~\cite{Ghosh:2021hrh}. 
Since the uncertainty quantification is not based on a robust statistical technique, attempting to reduce the uncertainty using AI may result simply in reducing our ad hoc estimate of the uncertainty, rather than the true uncertainty.

The best way to make concrete progress for such uncertainties is to develop a better understanding of their origin so that they can eventually be treated using a robust statistical model.

\section{Recommendations}


Below, we propose a list of guidelines in the context of interpretable UQ for our community to be better equipped to drive forward the progress in the field of AI-augmented HEP research. 
While some of these guidelines apply generally to good approaches in data-driven science research \cite{goodman2014ten}, we focus on recommendations pertinent to AI. 
\textbf{Overall, we emphasize that UQ for AI must become physically interpretable for usage in physics context.}

\begin{itemize}

\item Develop rigorous tests of local calibration. 

\item The HEP, Statistics, and AI communities should increase interaction to develop language that accurately reflects concepts relevant to both communities. Within 10 years, we should have a united lexicon for discussing and assessing interpretable uncertainties in AI for HEP.

\item The HEP-Stats-AI community should create benchmark data sets for rigorous testing and comparison of approaches to physically interpretable AI UQ for physics.

\item Funding agencies should endorse challenges and competitions within funding opportunity announcements to create and compare methods of UQ, including bias mitigation.


\item Develop and compare methods of feature visualisation and interpretability, explainability,  physics-awareness tools -- e.g., Local interpretable model-agnostic explanations (LIME) values, Shapley values, partial dependency plots, or Accumulated local effects (ALE) \cite{carvalho2019machine}.

\item Bring physically interpretable AI UQ to non-physics problems where appropriate (e.g., social science) to help improve risk assessment.

\item Common AI UQ methods should be embedded into deep learning software suites, similar to SKLearn, to enable widespread usage, testing, and comparison in the HEP community. 

\item AI adoption in HEP continues to lag behind latest methodologies. The HEP community should prioritize finding, testing, and comparing new methods generated by the AI community as soon as they emerge.

\end{itemize}
\section*{Acknowledgments}

BD acknowledges the support of the National Science Foundation under Grant No. AST-2009251. Any opinions, findings, and conclusions or recommendations expressed in this material are those of the author(s) and do not necessarily reflect the views of the National Science Foundation. AG is supported by The Department of Energy Office of Science. MK is supported by the US Department of Energy (DOE) under grant DE-AC02-76SF00515. Work at Argonne National Laboratory was supported by the U.S. Department of Energy, Office of High Energy Physics. Argonne, a U.S. Department of Energy Office of Science Laboratory, is operated by UChicago Argonne LLC under contract no. DE-AC02-06CH11357. N.R acknowledges the Laboratory Directed Research and Development (LDRD) funding from Argonne National Laboratory, provided by the Director, Office of Science, of the U.S. Department of Energy under Contract No. DE-AC02-06CH11357. 

\bibliographystyle{unsrt}
\bibliography{main.bib}

\end{document}